\documentclass[pra,aps,showpacs,floatfix,preprint]{revtex4-1}
\usepackage{color,colordvi,hyperref}
\usepackage{graphicx,amssymb,epsfig,amsmath,amsfonts}

\begin{document}

\title{Correlations of strongly interacting ultracold dipolar bosons in optical lattices}
\author{Budhaditya Chatterjee}
\email{bchat@iitk.ac.in}
\affiliation{Department of Physics, Indian Institute of Technology-Kanpur, Kanpur 208016, India}
\author{Marios C. Tsatsos}
\email{marios@ifsc.usp.br}
\affiliation{S\~ao Carlos Institute of Physics, University of S\~ao Paulo, PO Box 369, 13560-970, S\~ao Carlos, SP, Brazil}
\author{Axel U. J. Lode}
\email{axel.lode@univie.ac.at}
\affiliation{Wolfgang Pauli Institute c/o Faculty of Mathematics, University of Vienna, Oskar-Morgenstern Platz 1, 1090 Vienna, Austria}
\affiliation{Vienna Center for Quantum Science and Technology, Atominstitut, TU Wien, Stadionallee 2, 1020 Vienna, Austria}

\begin{abstract}
Strongly interacting dipolar bosons in optical lattices exhibit diverse quantum phases that are rich in physics. As the strength of the long-range boson-boson interaction increases, the system transitions across different phases: from a superfluid, through a Mott-insulator to a Tonks gas and, eventually, a crystal state. The signature of these phases and their transitions can be unequivocally identified by an experimentally detectable order parameter, recently described in \href{https://arxiv.org/abs/1708.07409}{arXiv:1708.07409}. Herein, we calculate the momentum distributions and the normalized Glauber correlation functions of dipolar bosons in a one-dimensional optical lattice in order to characterize all their phases. To understand the behavior of the correlations across the phase transitions, we first investigate the eigenfunctions and eigenvalues of the one-body reduced density matrix as the function of the  dipolar interaction strength. We then analyze the real- and momentum-space Glauber correlation functions, thereby gaining a spatially and momentum-resolved insight into the coherence properties of these quantum phases. We find an intriguing structure of non-local correlations that, independently of other system parameters, reveal the phase transitions of the system. In particular, spatial localization with synchronous momentum delocalization accompanies the formation of correlated islands in the density. Moreover, our study showcases that precise control of intersite correlations is possible through manipulation of the depth of the lattice, while intrasite correlations are influenced solely by changing the dipolar interaction strength.
\end{abstract}
\maketitle

\section{Introduction}
Ultracold atoms with dipole-dipole interactions have become a popular tool to simulate and understand the physics of long-range interacting systems ~\cite{baranov08,lahaye09}. The experimental realization of dipolar quantum gases has been achieved with atoms having permanent magnetic dipole moments, such as chromium~\cite{griesmaier05,beaufils08}, dysprosium~\cite{lu11} and erbium~\cite{aikawa12} as well as with polar molecules, for instance, potassium-rubidium~\cite{ni08} and cesium-rubidium \cite{nagrel14}. 
Owing to the long-range and anisotropic nature of dipole-dipole interactions,  novel quantum effects emerge not present in atoms with contact interactions. Prominent examples include the elongation of the condensate along the orientation of the dipole moments~\cite{yi01, santos00, goral00} and the exciting phenomenon of geometrical stabilization~\cite{lahaye09} of a dipolar Bose-Einstein condensate in traps of certain shapes, like extremely oblate ones~\cite{yi01,santos00,eberlein05,goral02a,koch08}. More recently, the fascinating formation of quantum droplets has been predicted and observed in dipolar condensates \cite{barbut2016,Chomaz2016,Waechtler2016,Macia2016}. 

In ultracold systems, the dimensionality is an experimental control-parameter and crucially important: lower-dimensional systems often produce a variety of effects not seen in three spatial dimensions. The occurrence of $p$-wave superfluidity in two-dimensional Fermi gases~\cite{bruun08, cooper09} provides such an example. 
Dipolar atoms in quasi-one-dimensional traps are more amenable experimentally, since the collisional instabilities arising from the head-to-tail alignment in two and three spatial dimensions are prevented ~\cite{koch08,ni10}. Unidimensional dipolar atoms have been predicted to exhibit Luttinger liquid-like behavior~\cite{arkhipov05,citro07,depalo08,pedri08} as well as anisotropic effects in curved and ring geometries~\cite{zollner11,zollner11pra,maik11}. Moreover, for very strong dipolar interactions a remarkable crystallization effect takes place where the dipolar atoms themselves form a crystal lattice structure irrespective of external confinements \cite{arkhipov05,astrakharchik08a,astrakharchik08b,Deuretzbacher2010,zollner11,chatterjee12,chatterjee17a}. 

Optical lattices often serve as a controllable toolbox to understand and simulate a large variety of condensed matter systems. For dipolar atoms, the additional existence of the long-range anisotropic interactions leads to a plethora of interesting quantum phases arising from the interplay of the kinetic energy,  the short and long-range interactions, each dominating different energy scales ~\cite{lahaye09}. Density waves~\cite{goral02, dalla06}, Haldane insulators~\cite{dalla06, deng11}, checkerboard patterns~\cite{goral02,menotti07} and Mott solids~\cite{zoller10} are some prominent examples of these  phases. 

In our study, we consider a system where four different phases are amalgamated: superfluid, Mott insulator \cite{Capello2007}, fermionized Tonks gas \cite{Petrov2000,Dunjko2001,Paredes2004,Deuretzbacher2007} and a crystal-like state \cite{Deuretzbacher2007,Deuretzbacher2010,zollner11} can each emerge in a finely tuned system of  few dipolar bosons in a multiple well trap. Superfluidity appears due to the bosonic nature of the particles combined with their weak interactions. A finite well depth together with somewhat stronger interactions breaks superfluidity and leads the system to behave as a so-called Mott-insulator \cite{jaksch98,greiner02}. For moderately strong  interactions, the system mimics the boson-to-fermion mapping, that is known to apply exactly at the infinite-strength limit of contact interactions \cite{girardeau60}. This is the Tonks-Girardeau limit and there the particles isolate themselves from their neighbors in order to avoid infinities in the interaction energy. Thus the bosonic density approaches that of its non-interacting fermionic counterpart. Nevertheless, the momentum density of the bosonic system is still distinct \cite{girardeau60}. We remark here that an exact Bose-Fermi map for long range dipolar interaction also exists, exploiting the divergence of the dipolar interaction at zero separation \cite{Deuretzbacher2010}. Last, for even higher interaction strengths, the long-range tail of the interaction dominates and leads to the formation of the so-called crystal phase \cite{arkhipov05,astrakharchik08a,astrakharchik08b,Deuretzbacher2010,zollner11,chatterjee12,chatterjee17a}. 

Herein we follow the same strategy as  Refs.~\cite{selim1,selim2,selim3} and theoretically investigate, the physics of a larger many-body system by studying in detail its few-body building blocks. To this end, we investigate a system of dipolar bosons in an optical lattice by studying the triple-well potential. Theoretically, dipolar atoms in triple-well traps have been explored using mean-field methods~\cite{peters12}, the extended-Bose-Hubbard model~\cite{lahaye10,anna13,fischer13,gallemi13,gallemi16,biedron18} and also the Multi-Configurational Time-Dependent Hartree (MCTDH) method~\cite{chatterjee12}. Notably, the mean-field methods and Bose-Hubbard model are unable to address very strong dipolar interactions. It is thus necessary to employ a general many-body approach for the cases where the strong dipolar interactions dictate the physics of the system.  

The MCTDH for bosons (MCTDHB) ~\cite{alon08} is such a general many-body method capable of addressing strong interaction regimes  \cite{cao17,Mistakidis15,Koutentakis17,Roy2017,bera18} and its implementation in the MCTDH-X software~\cite{ultracold,axel1,axel2} has been employed in Ref.~\cite{chatterjee17a} to establish an order parameter and an experimental method to classify and detect all  the quantum phases of dipolar atoms in optical lattices. 

In this paper we explore  the transitions across the above-mentioned four phases by varying the interaction strength and then analyze the  normalized Glauber correlation functions to gain  insight into the coherence properties for each phase. We show how the system can be brought to any of the desired phases and how the transitions are reflected in the correlation landscape, in the coordinate and momentum spaces.  We observe distinct  structural changes in the correlations that accompany these transitions, thereby clearly characterizing the phases and their transitions.  

We remark that the few-body finite size system studied here cannot exhibit true macroscopic phases. What we obtain are ground-states that are the finite-size precursor to the macroscopic thermodynamic phases \cite{Luhmann08}. For the sake of simplicity and as an analogy to the  thermodynamic systems, we still use the term ``phase" to refer to them.   

This paper is structured as follows: Sec.~\ref{sec:setup} introduces the Hamiltonian and quantities of our interest. Sec.~\ref{sec:NOs} discusses the eigenfunctions and eigenvalues of the reduced one-body density matrix for the emergent phases. Sec.~\ref{sec:gs} presents an analysis of the normalized Glauber correlation functions and Sec.~\ref{sec:outro} provides an outlook concluding our work.

\section{Model}\label{sec:setup}

Our system consists of $N$ polarized, dipolar bosons of mass $M$ in a one-dimensional optical lattice and is governed by the Hamiltonian
\begin{equation}
H= -\sum_{i=1}^{N}\frac{\hbar^{2}}{2M}\partial_{x_{i}}^{2}+\sum_{i=1}^{N}V_{ol}(x_{i})+\sum_{i<j}V_{int}(x_{i}-x_{j}).
\label{Eq.Ham}
\end{equation}
The one-body potential $V_{ol}$ represents a quasi-one-dimensional optical lattice potential, modeled as $V_{ol}= V\sin^{2}(\kappa x)$. Here, $V$ is the depth of the lattice and $\kappa$ its wave number. In order to confine the bosons to the desired number of sites, we impose a hard wall boundary condition at $x=\pm S\pi/2\kappa$ where $S$ is the number of lattice sites (for odd $S$). A strong transverse confinement of characteristic length $a_\perp$  ensures that the system is  quasi-one-dimensional by preventing excitations into the transverse direction.

We consider  a pure dipole-dipole interaction where the interaction potential  can be written as $V_{int}(x_i-x_j) = \frac{g_d}{|x_i - x_j|^3 + \alpha}$. For large separations $|x_i - x_j|\gg a_\perp$, the interaction potential varies as $V_{int}(x_i-x_j) \sim 1/|x_i - x_j|^3$. For small separations $|x_i - x_j|\lessapprox a_\perp$, the transverse confinement induces a short-scale interaction cutoff $\alpha \approx {a_\perp}^3$, thus regularizing the divergence at $x_i=x_j$ ~\cite{sinha07,Deuretzbacher2010,cai10}. 
Here $g_d$ is the dipolar interaction strength, given as $g_d = d^2_m /4\pi\epsilon_0$ for electric dipoles and as $g_d = d^2_m \mu_0 /4\pi$ for magnetic dipoles, $d_m$ being the dipole moment, $\epsilon_0$ the vacuum permittivity and $\mu_0$ the vacuum permeability.  In order to obtain universal dimensionless quantities we rescale Eq.~\eqref{Eq.Ham} in terms of the recoil energy $E_{R}=\hbar^{2}\kappa^{2}/2M$ effectively setting $\hbar=M=\kappa=1$. All quantities become dimensionless with the length given in units of $\kappa^{-1}$. 

We remark that  generally, the dipole-dipole interaction potential in 1D also includes a contact (Dirac delta) term  owing to the transverse confinement \cite{sinha07,Deuretzbacher2010,cai10}. However,  in order to examine the effect of the finite-range interaction alone, we set the contact interaction term equal to zero. In an experiment, this can be achieved via management of Feshbach resonances. 

We obtain the ground-states of the dimensionless Hamiltonian using the imaginary time-propagation of the MCTDHB equations of motion ~\cite{alon08}.  For the system studied, $M=7$ orbitals are sufficient to obtain the converged solutions.We use the values $S=3$, $V=8$, $N=6$ and $\alpha =0.05$ except otherwise specified.

\section{Natural orbitals, natural occupations and quantum phases}\label{sec:NOs}
While, in principle, the wavefunction contains all information about a quantum system, the aspects of correlations are better assessed through the reduced density matrix (RDM). Considering an $N$-particle state $|\Psi \rangle$, the $p^{th}$ order RDM is obtained by partially tracing the $N-p$ degrees of freedom \cite{sakmann08,Glauber63}
as: \begin{equation}
\rho^{(p)}=\mathrm{tr}_{p\dots N}|\Psi\rangle\langle\Psi|.\label{eq:rho2}
\end{equation}

For $p=1$, we obtain the $1^{st}$ order or one-body RDM 
\begin{eqnarray}
\rho^{(1)}(x,x')=\frac{1}{N} \langle \hat{\Psi}^{\dagger}(x) \hat{\Psi}(x') \rangle=\frac{1}{N}\int\Psi(x_1,\dots,x_N)\nonumber\\
\times \Psi^\ast(x'_1,\dots,x_N) dx_{2}\dots dx_{N}
\label{Eq.RDMp1}
\end{eqnarray}

where $\Psi\equiv\Psi(x_1,\dots,x_N)$ is the many-body 1D wavefunction in position space.
        
In the following, we analyze the one-body RDM  expanded in its eigenfunctions $\varphi_i$ as:  
 
\begin{equation}
\rho^{(1)}(x, x^{\prime})=\sum_i \lambda_i \varphi^*_i(x) \varphi_i(x^{\prime}).
\end{equation}
The eigenfunctions $\varphi_i$ are called natural orbitals and the corresponding eigenvalues $\lambda_i$ the natural occupations. Each $\lambda_i$ represents the population of the $i^{th}$ orbital. The spectral decomposition of the one-body RDM is particularly useful since it serves to define the Bose-Einstein condensation (BEC) in an interacting many-body system:
if the largest natural occupation is of  order of the number of particles $N$, the system is said to be condensed \cite{penrose56}. If there are more than one natural orbitals with populations of  order $N$ then the system is said to be fragmented \cite{nozieres82,Spekkens99,Mueller2006}.

\begin{figure}
\centering
\includegraphics[width=0.8\linewidth]{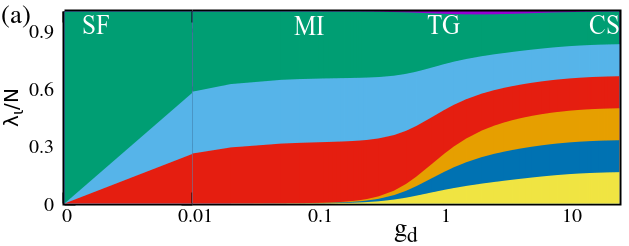}\\
\includegraphics[width=0.8\linewidth]{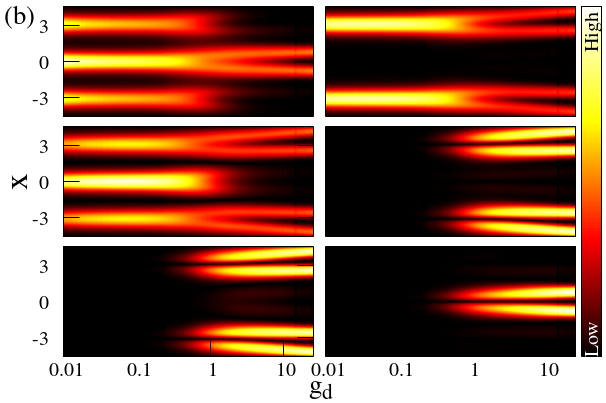}
\vspace{-5mm}
\caption{ Eigenfunctions and eigenvalues of the reduced one-body density matrix as  functions of the dipole-dipole interaction strength. (a) Normalized natural occupations $\lambda_l/N$ (plotted cumulatively) as functions of the interaction strength $g_d$. The region $0\leq g_d \leq 0.01$ is in linear scale. The region $g_d \geq 0.01$ is plotted in logarithmic scale. For weak interaction only one orbital is populated and the system is condensed. For moderately strong interactions, the first three orbitals $\phi_l,~l=1,2,3$ are occupied, reaching equal population $\lambda_l/N \approx 1/3$ and the system is in the Mott-insulating phase. For strong interactions, $6$ orbitals become equally populated, i.e. $\lambda_l/N \approx 1/6$ and the system is in the crystal phase.
 (b) Natural orbitals scaled by the natural occupations  -- $\lambda_i \times |\varphi_i|$ as a function of $g_d$ on log-scale. $\varphi_1$ which show the mean-field condensate contribution has a higher concentration in the central well. The combination of orbitals $\varphi_{1,2,3}$ is necessary to show the SF-MI transition. The orbitals $\varphi_{4,5,6}$ are occupied only for strong dipolar interactions and primarily contribute to the splitting in each well; this intrawell splitting is a hallmark of the breakdown of the Hubbard model and the emergence of the crystal state.
         }
\label{fig:frag}
\end{figure}
The natural orbitals and their populations are very important in characterizing the phases occurring for dipolar bosons in an optical lattice. The natural occupations directly relate to the emerging phases of strongly interacting dipolar bosons \cite{chatterjee17a,alon05,Roy2017}. Fig.~\ref{fig:frag} displays the evolution of the natural occupation as the function of interaction strength $g_d$.


In a conventional BEC with contact interactions, the interplay between the contact interactions, potential and kinetic energy determines the quantum properties ~\cite{greiner02,jaksch98}. For dipolar bosons, the additional long-range interactions make it a four-way competition between the kinetic, potential energy, the short-range and the long-range interactions that lead to the  existence of various phases at different energy scales \cite{chatterjee15, chatterjee17a}.
When the interaction is very small ($g_d\approx0$) the kinetic energy dominates, thus leading to a superfluid (SF) phase. Here the system forms a BEC since only the first natural orbital is macroscopically populated. Hence, in this region $\lambda_1 \approx N$. As $g_d$ increases the interaction energy starts to  dominate over the kinetic energy and the system arrives at the Mott-insulator (MI) phase  \cite{greiner02,bloch07,Capello2007,Roy2017}. In the MI phase, the bosons are localized inside each well and occupy $S$ (i.e. a number equal to the number of sites) orbitals equally, hence $\lambda_i=N/S$. At the same time tunneling between the lattice sites is strongly reduced \cite{chatterjee17a}. 
As  $g_d$  increases further, the long-range effect of the interaction now affects the many-body state and   the crystal phase is reached \cite{Deuretzbacher2007,Deuretzbacher2010,zollner11}. The already localized bosons in each well maximize their spatial separation due to the strong repulsion and the RDM shows maximal fragmentation with $N$ orbitals equally populated, i.e. $\lambda_i=1$ . 
Note that the formation of the crystal phase is driven only by the long-range interaction potential and is independent of the lattice potential. An interesting consequence of Bloch oscillations has been experimentally observed recently \cite{Meinert2017}.
 We remark here that the Bose-Hubbard model cannot describe the crystal phase since the dominant energy scale is the long-range interactions and not the lattice potential.  

The following schematic orbital picture  qualitatively illustrates  the basic mechanism of fragmentation and fermionization of six particles in a triple well. 
At the non-interacting limit, the ground state is in a single superposition $\phi_{GP}$ of the three modes ($\phi_l,\phi_c,\phi_r$) that are localized left, center and right, respectively. The existence of hard walls at the border of our potential breaks the translational invariance of the Hamiltonian and so the central well is slightly denser compared to the other two. Neglecting this asymmetry, the particle configurations (i.e., distributions of bosons across the wells) are degenerate energetically.
 The bosonic superposition $\phi_{GP}$ is coherent and the ground state (GS) is condensed and superfluid. In other words, a single orbital $\phi_{GP}=\dfrac{1}{\sqrt{3}}(\phi_l+\phi_c+\phi_r) $ is enough to describe the ground-state of the many-body system, as the product state \( \bigotimes_{i=1}^{N}\phi_{GP}(x_i)\equiv|N,0,0,\dots,0 \rangle\). 
Now consider a small but finite interaction parameter $g_d$. If the lattice sites are firmly separated from each other via sufficiently high barriers then the orbitals $\phi_l,\phi_c,\phi_r$ are not overlapping. In that case, the three-fold fragmented state  $\Psi=|\dfrac{N}{3},\dfrac{N}{3},\dfrac{N}{3},0\dots0\rangle$ is energetically favorable for any finite interaction strength  against  $\Psi_{GP}=|N,0,0,\dots,0 \rangle$ since the fragmented configuration minimizes the interaction energy (see also Ref.~\cite{Streltsov2004}). Hence, for large barrier heights $V$, a tiny interaction among the particles disrupts condensation in favor of fragmentation. The three-fold shape of the trap determines the number of significantly occupied orbitals to three. For a total number of $N=6$ particles, two particles occupy each lattice site of the trap.
For further growth of the interaction, each of the two particles in the same site starts to considerably repel its neighbor; the  orbitals now tend to avoid overlap in order to minimize the (otherwise increasing) interaction energy. This is the onset of fermionization, i.e. a density profile where bosons mimick a non-interacting fermionic density. For higher $g_d$, where the long-range tail of the interaction potential begins to dominate, we enter the crystal phase where the orbitals completely dissociate and each boson occupies a separate  orbital, giving rise to a six-fold fragmented state. Here the particle distribution and orbital shape are determined by the long-range interaction solely and not the lattice potential.

The above picture is certainly simplified since the interactions will change the shape of the natural orbitals and their occupations. For finite interactions there is a  substantial intermixing of the non-interacting orbitals, which is  parity preserving and thus a numerically exact solution must be sought.  However, the above line of thought illustrates the idea of the emergence of the different phases and their relation to fragmentation.

In Fig.~\ref{fig:frag}(b) we plot the evolution of the natural orbitals $\varphi_i(x)$ scaled by the natural occupations $\lambda_i$ with the interaction strength $g_d$. The orbital $\varphi_1$, lowest in energy, exhibits the mean-field condensate physics. For weak interactions, it shows the larger center-well population expected in the superfluid phase. However, it is unable to predict the equalizing of populations across the three wells at the MI and crystal phase. 
While $\varphi_1$ alone captures the mean-field perspective, all three lowest orbitals $\varphi_{1,2,3}$ are necessary to account for the SF-MI transition and understand the Bose-Hubbard picture. The population realignment to the outer wells for the MI state is carried primarily by $\varphi_2$. 
In other words, the system fragments with equal $\lambda_1,\lambda_2,\lambda_3$ across the three natural orbitals whose density maxima are located at the minima of each well. For strong $g_d$, $\varphi_2$ and $\varphi_3$ show a splitting of the outer maxima. The higher orbitals $\varphi_{4,5,6}$ contribute significantly only at the strongly interacting crystal regime. The primary feature of the latter is the enhancement of the intrawell (i.e. referring to the same well) splitting at strong interactions. Most importantly, the crystal phase is realized by the equal contribution of all $N$ natural orbitals \cite{chatterjee17a}.

\section{Correlation functions and quantum phases}\label{sec:gs}

We now discuss the many-body coherence and correlation properties of the system at the distinct phases. To this end, we assess the real and momentum space correlations in the state $|\Psi\rangle$, by studying the  $1^{st}$ and $2^{nd}$ order Glauber correlation functions \cite{Glauber63}.

\subsection{Spatial $1^{st}$ order correlations}

The $1^{st}$ order Glauber correlation function 
\begin{equation}
g^{(1)}(x^{\prime},x) = \frac{ \rho^{(1)}(x|x^{\prime})} {\sqrt{\rho(x)\rho(x^{\prime})}}
\end{equation} 
is constructed by normalizing the one-body RDM  [Eq.~\eqref{Eq.RDMp1}] to the respective one-body density. It measures the proximity of the many-body state to a mean-field state with $\rho(x,x^\prime)=\rho(x,x)=\rho(x^\prime,x')\equiv\rho(x)$ (i.e. uniform off-diagonals) \cite{sakmann08,Roy2017,Lode2017,axel1,axel2,Kronke15,Mistakidis18}. $g^{(1)}$ is generally, a complex quantity and is associated with phase coherence, which can be accessed through interference experiments \cite{Cirac1996,Narachewski1996,Javanainen1997,Naraschewski1999}. $g^{(1)}$ is bounded within $[0,1]$, with $|g^{(1)}|=1$ implying perfect coherence and $|g^{(1)}|=0$ a complete absence of it. By construction, the diagonal of $|g^{(1)}|$ is always equal to unity, so the coherence investigation needs to be done on the off-diagonal: $1^{st}$ order coherence between any two points $x,x^\prime$ denotes off-diagonal long-range order (ODLRO). 

\begin{figure}
\includegraphics[width=1\linewidth]{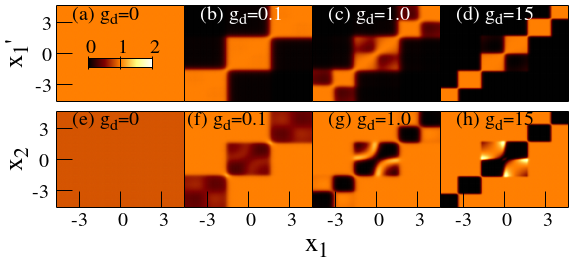}
\caption{(a)--(d): The $1^{st}$ order spatial correlation function $|g^{(1)} (x,x')|$ and (e)--(h): the $2^{nd}$ order correlation function $|g^{(2)} (x_1,x_2)|$, shown at all emergent phases. At $g_d=0$, the bosons are coherent, fact reflected in the uniform distributions: $|g^{(1)}|=1$ and $|g^{(2)}|=1$ everywhere. For the Mott insulator ($g_d=0.1$), the bosons are completely localized in each well and thus the long-range off-diagonal coherence is destroyed: first-order coherence is restricted to individual lattice sites while second-order coherence is characterized by particle bunching on the off-diagonal, i.e. $\vert g^{(2)}(x_1,x_2)\vert >1$ for $x\neq x'$ and anti-bunching within individual lattice sites, i.e. $\vert g^{(2)}(x_1,x_2)\vert <1$ for $x_1\sim x_2$. For stronger values of $g_d$ the intrawell first-order coherence decreases further. Also, the anti-bunching within sites is augmented ($\vert g^{(2)}(x_1,x_2)\vert \rightarrow 0$ for $x_1\sim x_2$) while the bunching between distinct sites disappears ($\vert g^{(2)}(x_1,x_2)\vert \rightarrow 1$) as the interaction strength increases. (All units shown are dimensionless.)}
\label{fig:g-x}
\end{figure}
        
Fig.~\ref{fig:g-x} displays $|g^{(1)} (x,x')|$ for various interaction strengths that correspond to the emerging distinct phases. For $g_d=0$ the bosons are delocalized over the whole lattice and are fully coherent. The many-body state is exactly described by a mean-field product of a single orbital $\phi_{GP}$. The complete coherence is reflected in the uniform distribution of $|g^{(1)} (x,x')|=1$ throughout the lattice. As the interaction is switched on, the ODLRO is disrupted, thus reflecting the particle localization at each well. In a Mott-insulator state at $g_d\approx0.1$, the bosons are completely localized and, as a consequence, coherence is strongly reduced; $|g^{(1)}|$ now shows a strong off-diagonal reduction. However, the localization is still partial, since inside each well the two residing atoms are delocalized and exhibit relative coherence. This intrasite delocalization and coherence is seen in the block-diagonal form of $|g^{(1)}|$ in Fig.~\ref{fig:g-x}(b). Here, each coherent block, occupying a space of roughly $2\times2$ square units, pertains to each lattice location.

As the interaction coupling increases, its short-range portion dominates \cite{interaction}, leading to fermionization of the bosons and a reduction of the intrawell coherence  \cite{Petrov2000,Dunjko2001,Paredes2004,Deuretzbacher2007}. At $g_d=1.0$, which falls in the Tonks gas regime, one can already see the off-diagonal contribution vanishing [Fig.~\ref{fig:g-x}(c)].

For a further rise of the interaction coupling, the long-range tail of the interaction begins to dictate the physics. The influence of the strong long-range tail of the interaction potential forces the localized bosons maximally apart, thus avoiding each other. As a  consequence, the coherence decreases even further.

Last, the system enters the crystal phase at $g_d=15$, where complete fragmentation characterizes the state. The bosons are entirely localized at each well, and the intrawell coherence is zero. The coherence blocks of $|g^1|$ are now centered at each boson location [Fig.~\ref{fig:g-x}(d)]. At the crystal phase, the bosons maximize their interparticle distance and do no longer necessarily correspond to the lattice spacing.
The absence of off-diagonal coherence reflects the complete real-space localization. 
Exceptionally, coherence marginally remains at a small region around the origin ($|g^{(1)}|$ being slightly above zero). 

\subsection{Spatial $2^{nd}$ order correlations}

The second order RDM $\rho^{(2)}(x_1,x_2,x'_1,x'_2)$ quantifies the correlations between two particles, with its diagonal kernel $\rho^{(2)}(x_1,x_2)$ representing the conditional probability of the simultaneous detection of a particle at  $x_1$ and another particle at $x_2$.

By normalizing it in terms of the respective one-body densities, we obtain the $2^{nd}$ order Glauber correlation function
\begin{equation}
g^{(2)}(x_1,x_2,x_{1}^{\prime},x_{2}^{\prime}) = \frac{ \rho^{(2)}(x_{1}, x_{2}|x_{1}^{\prime}, x_{2}^{\prime}) }{\sqrt{\rho(x_{1})\rho(x_{1}^{\prime})\rho(x_{2})\rho(x_{2}^{\prime}) }},
\end{equation}
which quantifies the $2^{nd}$ order coherence in the system. Hereafter, we use the diagonal $|g^{(2)}|\equiv|g^{(2)}(x_1,x_2)| = |g^{(2)}(x_1,x_2,x_1^\prime=x_1,x_2^\prime=x_2)|$.

For weak interactions ($g_d\approx0$) the state is at the SF phase yielding $|g^{(2)}|=1$, a fact that demonstrates $2^{nd}$ order coherence and the absence of any correlation in the measurement of the positions of any particle pair. In general, a SF phase shows $N^{th}$ order coherence \cite{sakmann08} and any $p$-particle detection probabilities are not correlated.
As the interaction increases, the diagonal of $|g^{(2)}|$ displays regions with anticorrelations $|g^{(2)}|< 1$ [Fig.~\ref{fig:g-x}(f)] thus demonstrating loss of coherence  and localization, or else \emph{anti-bunching}. In analogy to photon count statistics (see \cite{Emary2012} and references therein), we use here the term (anti-)bunching to denote (decreased) increased value of $g^{(2)}(x_1,x_2)$ -- as compared to unity. In the context of quantum optics, (anti)correlated emission events determine the detection probabilities and their distributions. Similarly, two bosons will bunch together and, hence, localize if $|g^{(2)}|>1$.
At the MI state, the anti-bunching block covers each lattice site, marking lack of coherence between the localized bosons in different sites.  For stronger interactions, the localized -- within the same site -- bosons lose coherence, forming the intrawell correlation holes [see Fig.\ref{fig:g-x}(g)]. These intrawell structures are the key signature of higher band effects that manifest both in the intermediate fermionization effect at $g_d=1$ [Fig.~\ref{fig:g-x}(g)] and the emergence of the crystal phase at $g_d=15$  [Fig.~\ref{fig:g-x}(h)]. While the diagonal pattern of $|g^{(2)}|$ shows no significant differences between the two cases, the off-diagonal background shows a strong reduction at $g_d=15$, displaying the emergence of extreme localization --  characteristic of the crystal phase.

\subsection{Momentum correlations}

We complement our investigations of the spatial correlations by analyzing the correlations $\tilde g^{(1,2)}(k,k')$ of the state $\Psi$ in the space of momenta $k$. Given that $\tilde g^{(p)}(k,\dots)$ is found from the RDM of the Fourier transformed $\tilde\Psi(k_1,\dots)$, and not as the Fourier transform of $\tilde g^{(p)}(x,\dots)$, it yields additional information about the structure of the state, not seen in $g^{(p)}(x,\dots)$ \cite{sakmann08}. In the following, we will drop the tilde for simplicity.

\subsubsection{$1^{st}$ order momentum correlations}
As previously done, we normalize the $1^{st}$ order RDM in momentum space with the local momentum densities and obtain the $1^{st}$ order momentum correlation function
                        
\begin{equation}
g^{(1)}(k^{\prime},k) = \frac{ \rho^{(1)}(k|k^{\prime}) }{
\sqrt{\rho(k)\rho(k^{\prime})}}.
\end{equation} 

\begin{figure}
\centering
\includegraphics[width=1\linewidth]{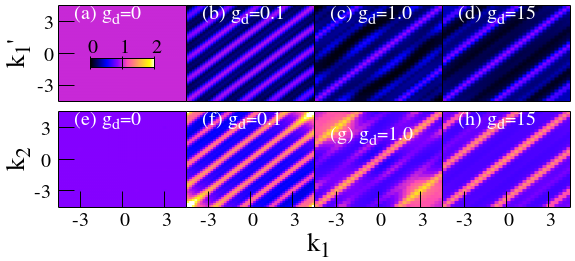}
\caption{(a)--(d): $1^{st}$ order momentum correlations and (e)--(h): $2^{nd}$ order momentum correlations for the distinct emergent phases of dipolar bosons in lattices. In the Mott-insulating state $g^{(1)}(k_{1}^{\prime},k_{1})$ shows a clean diagonal pattern of stripes with alternating strong $\vert g^{(1)} \vert \approx 1$ and weak $\vert g^{(1)} \vert \approx 0$ correlations. As the interaction strength is increased to $g_d=1.0$, a diagonal correlation hole (black regions with $\vert g^{(1)} \vert \approx 0$) develops. A striped pattern similar to the Mott-insulating phase, but with a different spacing  is seen in the crystal phase at $g_d=15$. The  $2^{nd}$ order momentum correlations $g^{(2)}(k_1,k_2)$ show similar tendencies but with their value lying in the interval $[1,2]$ thus signifying positive correlation.}
\label{fig:g-k}
\end{figure}

Fig.~\ref{fig:g-k} (top panel) shows  $g^{(1)}(k^{\prime},k)$ for the various emerging phases. At $g_d=0$, the system is fully coherent  and $g^{(1)}$ has  a constant value throughout the $k$ space corroborating our observations of the $x$ space correlations. As the interaction increases, $g^{(1)}(k^{\prime},k)$ shows an intricate diagonal pattern of alternating strong ($\vert g^{(1)} \vert \approx 1$) and weak ($\vert g^{(1)} \vert \approx 0$) correlations [Fig.~\ref{fig:g-k}(b)]. With stronger interaction ($g_d=1$), larger non-coherent regions form black stripes in Fig. \ref{fig:g-k}(c). As the crystal phase is approached at $g_d=15$, the diagonal non-coherent regions broaden, forming  an alternating  high-low pattern [Fig.~\ref{fig:g-k}(d)] but with a different spacing than Fig.~\ref{fig:g-k}(b).

\subsubsection{$2^{nd}$ order momentum correlations}

In Fig. \ref{fig:g-k} (bottom panel) we show the $2^{nd}$ order momentum correlation that offers insight into the $2^{nd}$ order coherence of the state at distinct phases. The superfluid regime $g_d=0$ [Fig.~\ref{fig:g-k}(e)] shows $2^{nd}$ order coherence everywhere in $k$-space. At interaction strength $g_d=0.1$ and the MI regime, a periodic bunching--anti-bunching  pattern emerges [Fig.~\ref{fig:g-k}(f)]. For stronger interactions the stripes in $g^{(2)}(k_1,k_2)$ become wider and lose contrast [Fig.~\ref{fig:g-k}(g--h)].



The form of the $k$-space correlations admits a simple interpretation. For weak interactions the gas is superfluid and this necessitates delocalization of the particles with complete $N$-order coherence. This is reflected in the constant value of $|g^{(2)}|=1$ that does not correlate any pair-measurement of momenta $k_1,k_2$. At the MI state, finding one particle with momentum $k_1$ indicates the (almost) certainty to find particle $2$ with $k_2=k_1+ n \delta_k$, $n$ integer, while no information is borne for any other $k_2$. Well after the fermionization limit has been crossed, the stripe-like pattern re-emerges [see  Fig.~\ref{fig:g-k}(h)], which can be interpreted as a fermionic diffraction pattern. Indeed, the existence of periodic patterns in $g^{(2)}(k_1,k_2)$ has been attributed to a phenomenon similar to Friedel oscillations in metals \cite{zollner11pra} where the (effective) fermion pair wavefunctions interfere. An examination of the structure of $g^{(2)}(k_1,k_2)$ reveals that the spacing of the stripes is $\delta_k\approx 2.3$ corresponding to a distance of $\delta_x\approx 1.36$ in real space, which equals the extension of each fermionized wavefunction (half lattice-site width). This periodic pattern is more clearly reflected in the correlations at very large $g_d$ [Fig.~\ref{fig:g-k}(h)] when in the crystal phase: there, the momentum distributions are expected to match the ones of interacting fermions \cite{Deuretzbacher2010}. 
Interestingly, the stripe-like periodic structure of $g^{(1)}$ and $g^{(2)}$ seen in the strongly interacting -- fermionized -- limit has been found for interacting fermions as well \cite{Brandt2017}.

\section{Conclusions}\label{sec:outro}

In this work, we have explored the many-body correlations of strongly interacting dipolar bosons in optical lattices. 
As the dipolar interaction strength increases, the bosons transition from a superfluid to a Mott-insulating and, eventually, for stronger dipolar interactions, to  a crystal phase. The phases are characterized by the natural orbitals and their populations. In the superfluid state, only one orbital is populated. In the Mott-insulating state the orbitals that are equally populated are as many as the sites in the lattice. Finally, for even strong dipolar interactions, the crystal phase emerges and the orbitals (equally populated) are as many as all the particles in the system. We thus verify  a fundamental connection between the natural occupations and the strength of the dipole-dipole interaction \cite{chatterjee17a}. Moreover, we examined the dependence of the natural orbitals on the interaction strength and showed how the densities in distinct quantum phases are built from them. 
Note that up to the value $g_d\approx 0.1$, all occupied natural orbitals topologically resemble the ones of the non-interacting system. Past  $g_d\approx 0.1$, the natural orbitals become topologically distinct.

As the interactions are increased, dipolar bosons in optical lattices cover the full range of correlation properties: while the weakly interacting superfluid is entirely uncorrelated, the long-range interaction-dominated crystal phase represents a strongly correlated state of a many-body system structured by the interparticle interactions and not the lattice potential. We showed how the normalized Glauber correlation functions of the ground-state undergo characteristic changes as  the interaction strength grows and the systems transitions across different phases. 

We comment on a fundamental difference between the SF$\rightarrow$MI and the MI$\rightarrow$CS transitions, which can be exploited for control and management of the system correlations. The former transition is extrinsic, implying that it can be induced via the one-body potential. The latter is intrinsic, i.e. it is only accessed via the two-body interaction potential. Hence,  precise control of the intersite (i.e. across different sites) correlations is possible through the manipulation of the depth of the lattice potential while the intrasite correlations can be controlled solely by the dipolar interaction strength. See Appendix \ref{A:depth} for details.

It is worth mentioning that, from Figs.~\ref{fig:g-x}, the relation $|g^{(1)}|+|g^{(2)}|\cong2$ seems to hold, for at least intermediate values of the interaction strength. This identity could be an expression of the Wick's theorem, that relates the value of higher-order correlation functions (propagators) to lower-order ones \cite{Dall2012}. Further investigations in that aspect could reveal useful connections.



Our current investigation of systems with dipole-dipole interactions provides a launchpad for further studies into the fundamental aspects of many-body correlations. A straightforward extension would be the calculation of higher-order correlation functions and also the investigation of lattices with incommensurate filling. Incommensurate systems are indeed drastically different from their commensurate counterparts \cite{brouzos10, Cinti16} and possessing a stronger sensitivity to the exact particle numbers. Thus significantly different ground-state properties are expected to be seen.
A more general study that includes contact interactions along with dipolar ones could reveal the interplay between short- and long-range interactions with adjustable strengths and promises to show new multiscale phases \cite{Pizzardo16}.  

\acknowledgments{BC gratefully acknowledges  financial support from the Department of Science and Technology, Government of India under the DST Inspire Faculty fellowship. MCT acknowledges financial support by the S\~ao Paulo state research foundation (FAPESP) and also CePOF/USP. AUJL acknowledges financial support by the Austrian Science Foundation (FWF) under grant No. F65 (SFB ``Complexity in PDEs''), and the Wiener Wissenschafts- und TechnologieFonds (WWTF) project No MA16-066 (``SEQUEX''). Computation time on the HPC2013 cluster of the IIT Kanpur and the Hazel Hen cluster of the HLRS in Stuttgart are gratefully acknowledged. }
\clearpage

\newpage
\appendix
\section{One-body reduced density matrix}
\label{A:RDM_x}
The reduced density matrix is important not only for its spectral decomposition but also from the perspective of its integral kernel $\rho^{(p)}(x_1,..x_p,x_1',..x_p')\equiv\langle x_1..x_p|\rho^{(p)}|x_1'..x_p'\rangle$. The latter relates to the correlation and coherence properties of the system and is used to construct the generalized $p^{th}$ order correlation functions \cite{sakmann08} 

\begin{equation}
g^{(p)}(x_1,..x_p,x_1',..x_p') = \frac{ \rho^{(p)}(x_{1}....x_{p}|x_{1}^{\prime}...x_{p}^{\prime}) }{
\sqrt{\prod_{i=1}^p\rho(x_{i})\rho(x_{i}^{\prime})}}.
\end{equation}

The $1^{st}$ order correlations ($p=1$) and hence the one-body aspects of the system are contained in the one-body RDM $\rho^{(1)}(x,x')$.
Its diagonal kernel $\rho(x)\equiv\rho^{(1)}(x,x)$ is real-valued and gives the one-particle density or equivalently the probability of finding a particle in position $x$ irrespective of all other $N-1$ positions. In contrast, the off-diagonal part is in general complex and thus not directly experimentally observable. 
Physically, it represents the overlap of a particle state at position $x$ and $x'$. 
                
The off-diagonal kernel of $\rho^{(1)}$ relates to the coherence properties of the system. 
For an infinite homogeneous system,  non-vanishing off-diagonals $\rho^{(1)}(x,x')$ as $|x-x'| \to \infty$ for all pairs $(x,x')$ imply off-diagonal long range order (ODLRO) and hence indicate  coherence \cite{Yang62}. For a finite spatially bounded system, there is no strict ODLRO. Instead, for finite systems coherence is established when the off-diagonal of $\rho^{(1)}(x,x')$ is simply non-vanishing (without further requirements).

\begin{figure}
\centering
\includegraphics[width=0.6\linewidth]{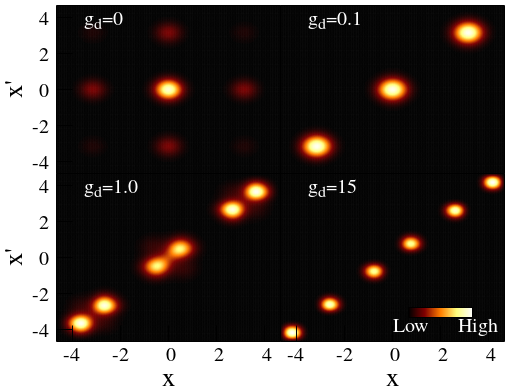}[h!]
\caption{ One-body reduced density matrix $\rho^{(1)}(x,x')$ for various interaction strengths. For $g_d=0$, the system is in a coherent superfluid state showing a checkerboard pattern with an intensity maximum at the origin. At $g_d=0.1$, the system is in the MI state. The intersite coherence is reduced and reflected in the lower density values at the  off-diagonals. At $g_d =1$, we are in the Tonks regime and the density maxima in each well begins to split. At $g_d=15$, we are deep in crystal phase and the complete splitting of the density maxima shows intrasite decoherence.}
\label{fig:rho1x}
\end{figure}

Fig.~\ref{fig:rho1x} shows the one-body reduced density matrix $\rho^{(1)}(x,x')$ for the various interaction strengths. In the limit of vanishing interaction, $g_d\rightarrow 0$ the bosons are condensed, and the state is superfluid. The comparatively high kinetic and potential energies lead to a greater population in the middle well. $\rho^{(1)}(x,x')$ thus shows a checkerboard pattern with a high value corresponding to the middle well and smaller values in the outer ones. As interaction is introduced, the bosons undergo localization in each well, whose degree increases with $g_d$ and is maximum at the MI phase for $g_d=0.1$. This localization reflects in $\rho^{(1)}(x,x')$ which shows three diagonal maxima corresponding to each well, meaning that the probability of finding a particle in $x$ and $x'$ is significant only within each of the lattice sites. As the repulsive interaction increases, the particles inside each site begin to separate and thus localize individually. 
At the crystal phase $g_d=15$, the bosons show complete intrasite localization with $\rho^{(1)}(x,x')$ showing $N=6$ separate diagonal maxima corresponding to each particle.

\section{Momentum reduced density matrix}
\subsection{One-particle distribution}

\label{A:RDM_k}
As in the real-space density, the single-particle momentum density matrix relates to the momentum coherence of the system. Physically the first order momentum density matrix represents the probability of finding a particle with momenta $k$ and $k'$. 

\begin{figure}
\centering
\includegraphics[width=0.8\linewidth]{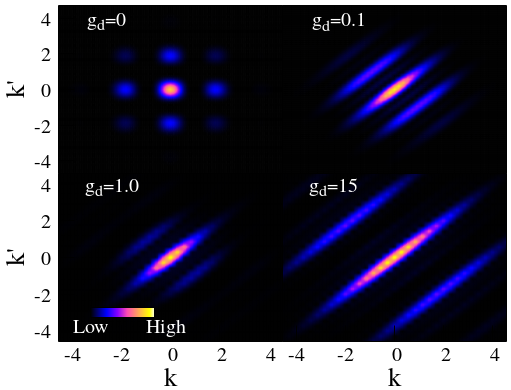}
\caption{ One-body reduced momentum density matrix $\rho^{(1)}(k,k')$ for various interaction. With interaction, the density shows diagonal stripes. At MI state $g_d=0.1$, there are three primary diagonal stripes along with fainter outer ones. At the crystal phase the stripes are spread outwards signifying distant momentum correlations. }
\label{fig:rho1k}
\end{figure}

In the uncorrelated limit, $g_d=0$, the momentum distribution is concentrated around $k=0$ showing a checkerboard pattern with secondary momentum coherence at the reciprocal lattice sites (Fig. \ref{fig:rho1k}). When the interaction in introduced the coherence centers stretch diagonally forming a striped diagonal pattern  with the principal diagonal having a maximum value. This reflects the localization in real space and, hence, delocalization in momentum space.

At the Mott-Insulator phase $g_d=0.1$, $\rho^{(1)}(k,k')$ shows three primary diagonal stripes pertaining to the three lattice sites.  The side-bands form from  stretching and joining the smaller coherence center at the reciprocal sites. There are also fainter bands at the edges formed from the stretching of the off-diagonal corners. Increasing interaction results initially, in a reduction of the distant stripes as can be seen at $g_d=1.0$. However, as the interaction is further increased, the distant bands reappear while the nearer ones diminish. 
One can understand this  behavior from the fact that there emerge two competing tendencies. The stronger repulsive interaction favors the enhancement of momentum correlations in the vicinity of $k=0$. However, as the interaction increases, the condensate localizes in real space, resulting in the spreading of the momentum correlations.

At the crystal phase $g_d=15$, we get three well-separated stripes, signifying distant momentum correlations.

\subsection{Two-particle distribution}
The  diagonal of the two-body momentum density (or, two-particle momentum distributions) gives the probability of finding one particle with momentum $k_1$ and another with momentum $k_2$. For $g_d \approx 0$ the $\rho(k_1,k_2)$ is concentrated near $k_1=k_2=0$ with small contributions at $(0,\pm \pi/2)$. In this regime, as discussed, only one orbital contributes and the momentum distribution essentially reflects the diffraction of this orbital on the lattice.

\begin{figure}
\centering
\includegraphics[width=0.8\linewidth]{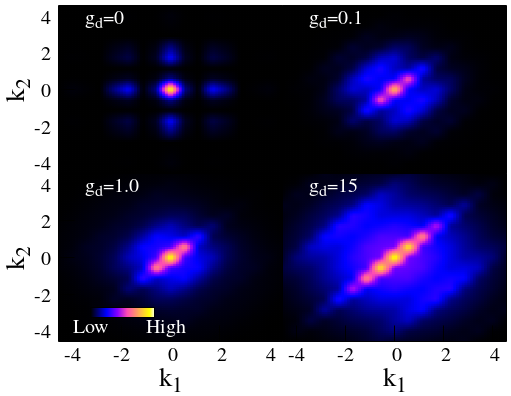}
\caption{The two-particle momentum distribution $\rho^{(2)}(k_1,k_2)$ for various interaction strengths. At $g_0=0$ we obtain a checkerboard pattern with peaked at the origin. In the MI state ($g_d=0.1$) there appear three peaks over a background cloud, which are connected to the threefold fragmented state. At the crystal state the momentum distributions of strongly interacting bosons become equal to these of interacting fermions \cite{Deuretzbacher2010}.}
\label{fig:den2dk}
\end{figure}

For increasing interaction ($g_d=0.1$) the discrete pattern now changes to a smeared distribution. The momentum distributions concentrate on the diagonal forming three lobes over continuous diagonal. This is contrasted to the two-body density that has a correlation hole at $x_1=x_2$. 
At $g_d=15$, the system is at the crystal state and the six-fold fragmentation leads to six maxima lobes at the diagonal, with further expansion of the off-diagonal contribution.

\section{Effect of lattice depth}
\label{A:depth} 
The two distinct phase transitions studied show a very different dependence with respect to the lattice depth $V$. The SF $\rightarrow$ MI transition shows a very strong dependence in $V$ while the MI $\rightarrow$ CS one is practically independent of $V$. In this section, we assess the $V$ dependence of the many-body correlations in the vicinity of each phase transition.

\begin{figure}
\centering
\includegraphics[width=0.8\linewidth]{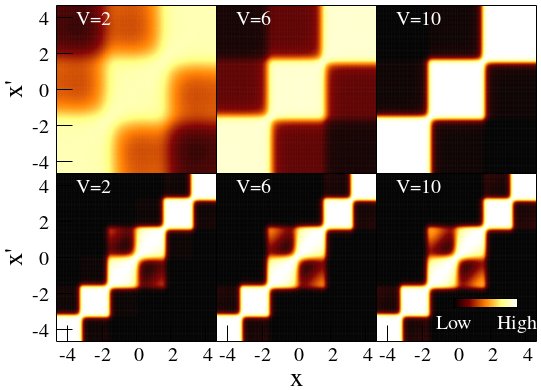}
\caption{The effect of the lattice depth. Shown is $|g^{(1)} (x,x')|$ for different lattice depths. Top panel: weak interaction ($g_d=0.02$). Large $V$ values cause loss of coherence. Thus by increasing $V$, one can induce the transition from SF to MI state. Bottom panel: strong interaction ($g_d=3.0$). Here, the value of the correlation function shows no change with increasing $V$. The MI to CS transition cannot be stimulated via $V$.}
\label{fig:V_d_g1x}
\end{figure}

                
Fig. \ref{fig:V_d_g1x} shows the first-order correlations for different $V$. At $g_d=0.02$ which is near the SF $\rightarrow$ MI, we see a stark dependence on $V$. For a small  depth $V=2$, the system is plainly superfluid and coherent. As the barrier is raised the ODLRO  strongly reduces as can be seen from the decrease of the off-diagonal values at $V=6$. At $V=10$  off-diagonal coherence is completely absent and a purely diagonal block structure emerges, indicating a MI phase. Thus, keeping $g_d$ fixed in the vicinity of the SF $\rightarrow$ MI transition one can induce the transition from the coherent SF phase to a non-coherent MI phase via $V$ alone.
Such transitions cannot be induced in the MI $\rightarrow$ CS crossover, which shows a very different character. The transition shows a minimal $V$ dependence that can be seen in the lower panel of Fig. \ref{fig:V_d_g1x}. 
Here, $|g^{(1)}|$ shows practically no dependence on $V$ in the vicinity of the phase transition $g_d=3.0$. The observations demonstrate the fundamental difference between the two phase transitions, as previously discussed. The SF $\rightarrow$ MI is an extrinsic transition since it can be assisted using the external (one-body) potential. The MI $\rightarrow$ CS one is, on the other hand,  a purely intrinsic transition, since it can be induced solely by the two-body dipolar interactions and cannot be triggered via the one-body potential.  Thus, the precise  control of intersite correlations can be achieved through the manipulation of the external potential while the intrasite can be controlled only through the dipole-dipole interaction coupling.


\end{document}